\documentclass[%
 preprint, 
 amsmath,amssymb,
 aps, physrev,
]{revtex4-2}
\usepackage[T1]{fontenc}
\usepackage[utf8]{inputenc}
\usepackage{physics}
\usepackage{comment}
\usepackage{tikz}
\usepackage{subcaption}
\usepackage{graphicx}
\usepackage{dcolumn}
\usepackage{bm}
\usepackage{hyperref}

\usepackage{lipsum}
\usepackage{xcolor}
\usepackage{mathrsfs}

\begin{document}

\title{Area–Information Trade-Offs in Acceleration Radiation from Atoms Falling into Black Holes}

%

\author{Yusef Maleki$^{1}$, Gustavo Valdivia-Mera$^{2,3}$,\\Carlos R. Ord\'{o}\~{n}ez,$^{2}$, Horacio E. Camblong$^{4}$, Marlan O. Scully$^{1,5}$}

\affiliation{\small $^{1}$  Institute for Quantum Science and Engineering, Texas A\&M University, College Station, TX 77843, USA
\\
$^{2}$ Department of Physics, University of Houston, Houston, TX 77204, USA
\\
$^{3}$ Texas Center for Superconductivity, University of Houston, Houston, TX 77204, USA
\\
$^{4}$ Department of Physics and Astronomy, University of San Francisco, San Francisco, California 94117-1080, USA
\\
$^{5}$ Princeton University, Princeton, NJ 08544, USA
}

\date{\today}\vspace{5mm}
\begin{abstract}
We develop a geometric theory of information processing in the
Horizon-brightened acceleration radiation (HBAR) channel, in which the
radiative horizon-area change provides an entropy budget for the information
carried by the radiation field. Building on the quantum-optical description
of atom--field interactions near the horizon and the resulting HBAR
thermodynamic correspondence, we derive area-cost laws in the
near-steady, thermally saturated regime. The accessible classical information
and the mutual information generated between the radiation field and its
environment are bounded by the associated radiative horizon-area budget.
Reliability is incorporated through Fano's inequality, which translates a
prescribed decoding error probability into an area requirement. We further
derive Fisher-information speed limits that constrain the statistical
evolution of the radiation field and place a lower bound on the duration
required for correlation generation. Together, these results establish a bits-per-area
principle linking black-hole thermodynamics, information geometry, and quantum
information in the HBAR framework.
\end{abstract}

\maketitle
\clearpage

\begingroup
\setlength{\parskip}{0pt}
\setlength{\parindent}{0pt}
\normalsize
\renewcommand{\baselinestretch}{1.4}\selectfont
\tableofcontents
\endgroup

\clearpage

\section{Introduction}

The interplay among information theory, thermodynamics, and gravitation
constitutes one of the most significant interfaces in theoretical physics,
with black holes occupying a central place in this setting
\cite{peres2004quantum,misner2009john,zych2011quantum,
bekenstein2004black,maleki2024cosmic}. The seminal works of Bekenstein
and Hawking established that a black hole is a thermodynamic system endowed
with an entropy proportional to the area of its event horizon and a
temperature determined by its surface gravity
\cite{Bekenstein:1972tm,bekenstein1973black,hawking1974black,
hawking1975particle,bekenstein2020black}. This remarkable connection
between geometry and thermodynamics naturally invites an
information-theoretic interpretation in which spacetime geometry constrains
the storage, transfer, and distribution of information.

The Unruh effect
\cite{fulling1973nonuniqueness,davies1975scalar,unruh1976notes} and Hawking
radiation~\cite{hawking1975particle} further show that the particle content of
a quantum field depends on the observer and on the presence of horizons,
giving rise to thermal phenomena in otherwise vacuum configurations. In this
context, the Horizon-brightened acceleration radiation (HBAR) process,
introduced in Ref.~\cite{scully2018quantum}, provides a quantum-optical
framework for exploring these effects~\cite{scully1966quantum}. Freely
falling two-level atoms interact weakly with a quantum scalar field and may
become excited while emitting scalar quanta because of their relative
acceleration with respect to the field modes selected by the
cavity~\cite{PhysRevLett.91.243004}. The resulting radiation-field density
matrix approaches a thermal steady state at the Hawking temperature, with the
detailed-balance relation governed by the near-horizon conformal
quantum-mechanical structure
\cite{camblong2005black,Camblong2005-SCBH,camblong2020near,
Ordonez2025QAspST}. This construction leads, in turn, to a formal
thermodynamic correspondence between the radiation field resulting from the
HBAR process and Kerr horizon
thermodynamics~\cite{Ordonez2025QAspST}.

While the HBAR entropy, related detector-response phenomena, and their
connections with black-hole thermodynamics have been explored in a variety of
settings
\cite{Ordonez2025QAspST,sen2022equivalence,jana2024atom,jana2025inverse,
chakraborty2019detector,das2024horizon,das2025derivative,ovgun2026hbar},
a geometric information-theoretic account of how the horizon area constrains
information processing in this framework has remained undeveloped. In
particular, two questions arise naturally: how much accessible classical
information can be supported by the acceleration radiation, and how does the
associated radiative horizon-area budget constrain the correlations generated
between the radiation field and its environment? The present work addresses
these questions by developing a unified geometric and information-theoretic
framework for the HBAR channel, in which the radiative contribution to the
horizon-area change acts as the geometric resource underlying a set of
area-cost laws.

More specifically, we first use Spohn's theorem to place the HBAR entropy
balance within the framework of irreversible quantum thermodynamics and to
distinguish the general nonnegative entropy production from its near-steady,
thermally saturated limit~\cite{Spohn1,Spohn2}. In this regime, the radiation
entropy, corotating energy, and radiative horizon-area change obey a common
balance relation. Combining this relation with the Holevo bound yields an
area-cost law for accessible classical information, while the Araki--Lieb
inequality gives the corresponding geometric bound on
radiation--environment mutual information
\cite{Holevo1973,araki1970entropy}. Decoding reliability is incorporated
through Fano's inequality~\cite{Fano1961,HanVerdu1994}. We then use the
dynamical Fisher information of the radiation-field occupation-number
distribution to relate its statistical evolution to the area and
corotating-energy fluxes and to derive a quantum speed limit on the duration
required for correlation generation
\cite{nicholson2020time,ito2020stochastic,maleki2020speed}. Together, these results
establish a bits-per-area principle linking black-hole thermodynamics,
information geometry, and quantum information in the HBAR framework.

The article is organized as follows. Section~\ref{sec:HBAR-Kerr} develops the
HBAR process in Kerr geometry, including the near-horizon conformal
quantum-mechanical structure, the atom--field interaction, the emission and
absorption rates, and the resulting thermal radiation-field density matrix.
Section~\ref{sec:HBAR-thermo} examines the thermodynamic content of this
density matrix and establishes its formal correspondence with Kerr horizon
thermodynamics. In Sec.~\ref{sec:entropy-area-Spohn}, the entropy balance is
reexamined from the perspective of Spohn's theorem, identifying the
near-steady saturated regime in which the entropy--energy--area relation
holds. Section~\ref{sec:bits_area} derives the area-cost law for accessible
classical information and its Landauer-type energetic interpretation, while
Sec.~\ref{sec:mi_area} develops the corresponding bounds for mutual
information and incorporates decoding reliability through Fano's inequality.
Section~\ref{sec:fisher_area} formulates the Fisher-area speed limits and the
corresponding lower bound on the duration required for correlation generation. Finally,
Appendix~\ref{sec:area_monogamy} presents the area-bounded sharing constraint
for bipartitions of the degrees of freedom complementary to the outgoing
radiation.

A remark on units is in order. In Secs.~\ref{sec:HBAR-Kerr} and
\ref{sec:HBAR-thermo}, we work in natural units,
\(c=\hbar=k_B=G=1\). The fundamental constants are restored beginning in
Sec.~\ref{sec:entropy-area-Spohn}, where the entropy, energy, area, and
information bounds are expressed in physical units.

\section{HBAR in Kerr geometry}
\label{sec:HBAR-Kerr}

This section develops the HBAR process in Kerr geometry. We begin with the
geometric and near-horizon field-theoretic ingredients required to formulate
the interaction between freely falling atoms and the scalar field, and then
analyze the resulting radiative dynamics at the level of transition rates and
the radiation-field density matrix.

\subsection{Kerr geometry}

The Kerr metric describes an exact, stationary, and axisymmetric vacuum solution
of the Einstein field equations in four dimensions for a rotating black hole
with mass \(M\) and angular momentum \(J\). Here, we follow the spacetime
conventions of Ref.~\cite{MTW-gravitation} and, unless otherwise stated, work
in natural units \(c=\hbar=G=k_B=1\). Then, in Boyer--Lindquist coordinates
\((t,r,\theta,\varphi)\), the line element takes the standard compact
form~\cite{MTW-gravitation,Ordonez2025QAspST}
\begin{equation}
    ds^2
    =
    -\frac{\Delta}{\rho^2}
    \left(dt-a\sin^2\theta\,d\varphi\right)^2
    +\frac{\rho^2}{\Delta}\,dr^2
    +\rho^2\,d\theta^2
    +\frac{\sin^2\theta}{\rho^2}
    \left[(r^2+a^2)d\varphi-a\,dt\right]^2,
    \label{kerr_metric1}
\end{equation}
where the Kerr parameter \(a=J/M\) represents the angular momentum per unit
mass, and
\begin{equation}
    \Delta=r^2-2Mr+a^2,
    \qquad
    \rho^2=r^2+a^2\cos^2\theta.
\end{equation}
The near-horizon analysis considered here also extends to a Kerr--Newman
background with electric charge \(Q\). In this case, the line element retains
the form of Eq.~\eqref{kerr_metric1}, with \(\Delta\) replaced by
\(\Delta_Q=r^2-2Mr+a^2+Q^2\).

The stationary and axisymmetric character of the metric~\eqref{kerr_metric1}
gives rise to two commuting Killing vector fields,
\begin{equation}
    \xi^{(t)} = \partial_t, 
    \qquad 
    \xi^{(\varphi)} = \partial_\varphi,
    \label{kill_vec}
\end{equation}
associated respectively with time translation symmetry and rotational
invariance about the symmetry axis. 

The locations of the horizons for a metric of the form~\eqref{kerr_metric1}
are determined by the vanishing of the radial component of the inverse metric,
\(g^{rr}=0\). Since this condition is equivalent to \(\Delta=0\), the two real
roots characterize the inner \((-)\) and outer \((+)\) horizons,
\begin{equation}
    r_\pm = M \pm \sqrt{M^2 - a^2}.
    \label{inotho}
\end{equation}
In Eq.~\eqref{inotho}, the outer root \(r_+\) corresponds to the event horizon,
provided the nonextremal condition \(M^2>a^2\) holds.

The following three quantities play a central role in the geometric and
thermodynamic characterization of the horizon: the area \(A\), the angular
velocity \(\Omega_H\), and the surface gravity \(\kappa\). The horizon area is
given by
\begin{equation}
    A 
    = \int \sqrt{g_{\theta\theta}(r_+, \theta) \, g_{\varphi\varphi}(r_+, \theta)} 
      \, d\theta \, d\varphi
    = 4 \pi (r_+^2 + a^2).
\end{equation}
The horizon-generating Killing vector is the linear combination
\begin{equation}
    \xi = \xi^{(t)} + \Omega_H \, \xi^{(\varphi)} ,
\end{equation}
which becomes null on the outer horizon \(\mathcal H_+\). The angular velocity
of the horizon, obtained from this null condition, is
\begin{equation}
    \Omega_H 
    = \lim_{r \to r_+} \qty(-\frac{g_{t\varphi}}{g_{\varphi\varphi}}) 
    = \frac{a}{r_+^2 + a^2} 
    = \frac{a}{2 M r_+}.
    \label{angvelhrp}
\end{equation}
The surface gravity, defined geometrically from the same horizon-generating
Killing vector \(\xi\), takes the compact form
\begin{equation}
    \kappa 
    = \frac{\Delta'_+}{2 (r_+^2 + a^2)},
    \label{surfgrv}
\end{equation}
where \(\Delta'_+ \equiv \left. d\Delta / dr \right|_{r = r_+}\). This
construction describes the relevant physics in a frame corotating with the
black hole and geometrically identifies \(\mathcal H_+\) as a Killing horizon.

\subsection{Scalar-field quantization and near-horizon conformal dynamics}
\label{sec:Quantization-NH-CQM}

A massless scalar field \(\Phi\) propagating in a curved spacetime with metric
\(g_{\mu\nu}\) satisfies the covariant Klein--Gordon equation
\begin{equation}
    \Box\Phi 
    \equiv 
    \frac{1}{\sqrt{-g}} \, \partial_\mu \!\left( \sqrt{-g} \, g^{\mu\nu} \partial_\nu \Phi \right) = 0.
    \label{eq:Klein_Gordon_basic}
\end{equation}

The corresponding quantization is enforced by the usual canonical Hamiltonian
rules, whereby the classical field and its conjugate momentum are promoted to
quantum operators satisfying the canonical commutation relations. This
quantization procedure is then reduced to the problem of finding a complete
set of solutions
\(\bigl\{ \phi_{\mathbf{s}}(t,\mathbf{r}), \phi^{*}_{\mathbf{s}}(t,\mathbf{r}) \bigr\}\)
of the classical equation~\eqref{eq:Klein_Gordon_basic} and expanding the
quantum field operator \(\hat{\Phi}\) as 
\begin{equation}
    \hat{\Phi}(t, \mathbf{r}) 
    = \sum_{\mathbf{s}} \bigl[
    \hat a_{\mathbf{s}} \phi_{\mathbf{s}}(t,\mathbf{r})
    + \mathrm{H.c.} \bigr],
    \label{eq:field_expansion_2}
\end{equation}
where \(\mathrm{H.c.}\) denotes the Hermitian conjugate and
\(\mathbf r=(r,\theta,\varphi)\) denotes the spatial Boyer--Lindquist
coordinates. The mode label \(\mathbf{s}=(\omega,l,m)\) comprises the
frequency \(\omega\), the spheroidal angular number \(l\), and the azimuthal
number \(m\). They are also assumed to
satisfy the orthonormality conditions
\((\phi_{\mathbf{s}},  \phi_{\mathbf{s}'} ) 
= - (\phi^{*}_{\mathbf{s}},  \phi^{*}_{\mathbf{s}'} ) 
= \delta_{ {\mathbf{s}}, {\mathbf{s}'} }\) and 
\begin{equation}
    (\phi^{*}_{\mathbf{s}},  \phi_{\mathbf{s}'} ) 
 = (\phi_{\mathbf{s}},  \phi^{*}_{\mathbf{s}'} ) 
   = 0,
\end{equation}
with respect to the standard inner product~\cite{birrell-davies,Crispino-et-al_2008},
chosen consistently with Eq.~\eqref{eq:Klein_Gordon_basic}.

The modes required for the expansion~\eqref{eq:field_expansion_2} can be
derived by using the stationary and axisymmetric character of the Kerr
geometry, encoded in the Killing vectors \(\partial_t\) and
\(\partial_\varphi\) [see Eq.~\eqref{kill_vec}]. Separation of variables then
gives the mode decomposition
\begin{equation}
    \phi_{\mathbf{s}}(t, r, \theta, \varphi)
    = R_{\mathbf{s}}(r) \, S_{\mathbf{s}}(\theta) \,
      e^{i m \varphi} e^{-i \omega t}.
    \label{eqfm12}
\end{equation}
Substitution of the mode decomposition~\eqref{eqfm12} into
Eq.~\eqref{eq:Klein_Gordon_basic} yields angular and radial equations linked
by a common separation constant.

A near-horizon analysis reveals that the relevant physics is provided by
conformal quantum mechanics (CQM), as established in
Refs.~\cite{camblong2005black,Camblong2005-SCBH} and later applied to the
scalar field in the HBAR process in
Refs.~\cite{Ordonez2025QAspST,camblong2020near}. Specifically, in the
near-horizon limit, denoted by the symbol \(\overset{(\mathcal{H})}{\sim}\) and
implemented through the expansion \(r=r_+ + x\) with \(x\ll r_+\), the dynamics
reduces to an effective radial equation depending only on the frequency
\(\omega\) and azimuthal number \(m\). Introducing the reduced radial function
\(u(x)\) through the redefinition \(R(x)\propto x^{-1/2}u(x)\), the
near-horizon limit of the Klein--Gordon equation becomes
\begin{equation}
    \frac{d^2 u(x)}{dx^2} + \frac{\lambda}{x^2} \left[ 1 + \mathcal{O}(x) \right] u(x) = 0,
    \label{eq:near_horizon}
\end{equation}
where the coupling parameter \(\lambda\) is given by
\begin{equation}
    \lambda = \frac{1}{4} + \Theta^2,
    \qquad
    \Theta = \frac{\tilde{\omega}}{2 \kappa},
    \qquad
    \tilde{\omega} = \omega - m \Omega_H .
\end{equation}
Here \(\kappa\) and \(\Omega_H\) denote, respectively, the surface gravity and
angular velocity of the horizon [see Eqs.~\eqref{surfgrv} and
\eqref{angvelhrp}], and \(\tilde{\omega}\) is the frequency associated with the
frame corotating with the horizon. Since \(\tilde{\omega}\) can be negative for
superradiant modes in Kerr, the thermal behavior discussed below is understood
for outgoing modes with positive corotating frequency, \(\tilde{\omega}>0\),
following the treatment of Ref.~\cite{Ordonez2025QAspST}.

The effective Hamiltonian associated with the leading-order part of
Eq.~\eqref{eq:near_horizon} can then be written as
\(H_{\rm CQM}=p_x^2+V_{\text{eff}}(x)\), where
\(V_{\text{eff}}(x)=-\lambda/x^2\). This system is classically scale invariant
and defines a one-dimensional CQM model governed by an inverse-square
potential
\cite{camblong2005black,Camblong2005-SCBH,
Ordonez2025QAspST,camblong2020near}. At leading order, after neglecting the
subleading \(\mathcal O(x)\) corrections, Eq.~\eqref{eq:near_horizon} admits
the basic power-law solutions
\begin{equation}
    u(x) = x^{1/2 \pm i \Theta},
\end{equation}
which correspond to outgoing and ingoing wave modes, respectively. These
solutions display a logarithmic-phase singularity characteristic of
scale-invariant systems. Restoring the time dependence, the corresponding
near-horizon field modes take the form
\begin{equation}
    \phi_{\mathbf{s}}(t,r,\theta,\varphi)
    \overset{(\mathcal{H})}{\sim}
    \phi^{\pm(\mathrm{CQM})}_{\mathbf{s}}
    \propto x^{\pm i \Theta} S_{\mathbf{s}}(\theta)
    \, e^{i m \tilde{\varphi}}
    \, e^{-i \tilde{\omega} t},
    \label{eq:CQM_modes}
\end{equation}
where \(\tilde{\varphi}=\varphi-\Omega_H t\) is the azimuthal coordinate in the
frame corotating with the horizon, and \(\tilde{\omega}\) again denotes the
corotating frequency.

Finally, we briefly summarize the near-horizon form of timelike geodesics in
the Kerr background, which describe the trajectories of freely falling atoms in
locally inertial frames. These trajectories are essential for computing the
emission and absorption probabilities of scalar quanta within the near-horizon
approximation:
\begin{align}
    \tau &\overset{(\mathcal{H})}{\sim}
     -k\,x + \mathcal{O}(x^2) + \text{const.},
    \label{nh_rau} \\
    t &\overset{(\mathcal{H})}{\sim} -\frac{1}{2\kappa}\ln(x) - C\,x + \mathcal{O}(x^2),
    \label{nh_t} \\
    \tilde{\varphi} &\overset{(\mathcal{H})}{\sim} \alpha\,x + \mathcal{O}(x^2),
    \label{nh_varphi}
\end{align}
where the constants \(k\), \(C\), and \(\alpha\) arise from the near-horizon
expansion of the geodesic equations and depend on the conserved geodesic data
and the geometric parameters of the Kerr metric~\eqref{kerr_metric1}. Their
explicit forms are given in Ref.~\cite{Ordonez2025QAspST} and need not be
reproduced here.

\subsection{Atom--field interaction in the optical-cavity setup}

The physical system under consideration consists of a quantum scalar field and
a freely falling cloud of two-level atoms that interact weakly along their
worldlines. The free-fall condition determines the worldlines to be geodesics
in the background geometry of a rotating Kerr black hole. This setup corresponds to a thought experiment in which the atoms are injected
randomly along their free-fall trajectories, while the scalar field is subject
to boundary conditions analogous to fixed mirrors in laboratory optics. These
boundary conditions implement the optical-cavity model and select a
Boulware-like vacuum~\cite{boulware1975quantum}, which is assumed for the
transition probabilities leading to the acceleration radiation. The original version of
this setup was proposed for the Schwarzschild geometry in
Ref.~\cite{scully2018quantum}, where the HBAR process was first introduced.
The relevant physics of this framework was subsequently shown to be described
by near-horizon CQM and extended to include rotating black holes in
Refs.~\cite{camblong2020near,Ordonez2025QAspST}.

The composite physical system includes both a real scalar field and a cloud of
atoms. First, the scalar field operator \(\hat{\Phi}(t,\mathbf r)\) is defined
on the Kerr spacetime as described in Sec.~\ref{sec:Quantization-NH-CQM}. The choice of the required Boulware-like vacuum state \(\ket{0_B}\) is
technically subtle due to the presence of superradiant
modes~\cite{starobinskii1973amplification,frolov2012black}. As discussed in
Ref.~\cite{Ordonez2025QAspST}, this difficulty can be circumvented by
introducing a reflecting boundary that excludes the asymptotic region, thereby
ensuring a well-defined quantization. This corresponds physically to the
optical-cavity model of the gedanken experiment described above. The resulting radiation field can be regarded as consisting of scalar
quanta, according to the quantization of Eq.~\eqref{eq:field_expansion_2}. This
scalar-field description is standard in quantum field theory~\cite{birrell-davies}
and quantum-optical~\cite{PhysRevLett.91.243004} treatments, as this
simplification captures the essence of the most relevant spacetime effects.
The extension to higher-spin fields is possible~\cite{birrell-davies,Crispino-et-al_2008},
though it makes the formalism more complicated.

The second subsystem consists of a dilute ensemble of atoms modeled as
identical two-level systems. In the present setup, the atoms are injected at
random times from an initial radius \(r_{\rm in}\gg r_+\) and subsequently
follow freely falling geodesic trajectories toward the outer horizon. During their free-fall motion, the atoms interact with the quantum
field. Although the field is initially prepared in its vacuum state, the
freely falling atoms are not comoving with the field modes selected by the
cavity. Because of their relative acceleration with respect to the field modes,
transition channels involving the emission and absorption of scalar quanta
become possible.

The weak interaction between each atom and the spin-zero field is modeled by
the monopole analog of a dipole coupling for a spin-one field. In the
interaction picture, the interaction Hamiltonian is~\cite{Ordonez2025QAspST}
\begin{equation}
    \hat V_I(\tau)
    =
    g\,\hat{\Phi}\bigl(t(\tau),\mathbf r(\tau)\bigr)
    \left(
        \hat\sigma_-e^{-i\nu\tau}
        +\mathrm{H.c.}
    \right),
    \label{intvi}
\end{equation}
where \(g\) is the atom--field coupling constant, \(\hat\sigma_-\) is the
atomic lowering operator, \(\nu\) is the atomic transition frequency, and
\(\tau\) is the proper time along the atom's worldline. The corresponding
raising operator is \(\hat\sigma_+=\hat\sigma_-^\dagger\).

Considering the initial state of the atom-field system as \(\ket{0_B, b}\),
where \(\ket{0_B}\) is the selected Boulware-like vacuum and
\(\ket{b}\) is the atomic ground state, the interaction~\eqref{intvi} allows a freely falling atom to become
excited and simultaneously emit a scalar quantum via the operator
\(\hat a^\dagger_{\mathbf{s}}\hat\sigma_{+}\)~\cite{PhysRevLett.91.243004}.
In the weak-coupling regime, first-order perturbation theory applies, yielding
the transition probability to the final state
\(\ket{1_{\mathbf{s}}, a}\), where \(\ket{1_{\mathbf{s}}}\) denotes a
one-particle field excitation and \(\ket{a}\) is the excited atomic state:
\begin{equation}
    P_{\text{e},\mathbf{s}} 
    = \left|\int d\tau\, 
    \bra{1_{\mathbf{s}}, a} \hat V_I(\tau) \ket{0_B, b}\right|^2
    = g^2 \left|\int d\tau\, 
    \phi^*_{\mathbf{s}}(t(\tau), \mathbf{r}(\tau))\, e^{i\nu\tau} \right|^2.
    \label{probem1}
\end{equation}

Similarly, another freely falling atom may absorb the emitted scalar quantum,
driving the transition \(\ket{1_{\mathbf{s}}, b}\to\ket{0_B, a}\) through the
operator \(\hat a_{\mathbf{s}}\hat\sigma_{+}\). The corresponding
absorption probability is
\begin{equation}
    P_{\text{a},\mathbf{s}} 
    = \left|\int d\tau\, 
    \bra{0_B, a} \hat V_I(\tau) \ket{1_{\mathbf{s}}, b}\right|^2
    = g^2 \left|\int d\tau\, 
    \phi_{\mathbf{s}}(t(\tau), \mathbf{r}(\tau))\, e^{i\nu\tau} \right|^2.
\end{equation}

\subsection{Emission and absorption rates}
\label{emabprb1}

The emission and absorption probabilities of scalar quanta by freely falling
atoms near the outer horizon can be obtained from the near-horizon analysis.
This requires the asymptotic behavior of the coordinates \((t,\tilde\varphi)\)
and the proper time \(\tau\), as given in
Eqs.~\eqref{nh_rau}--\eqref{nh_varphi}, together with the CQM modes in
Eq.~\eqref{eq:CQM_modes}. The outgoing \((+)\) mode corresponds to quanta propagating away from the
horizon toward increasing \(r\), while the ingoing \((-)\) mode corresponds
to quanta falling inward toward the outer horizon.

To characterize these processes, it is convenient to consider the
corresponding emission and absorption rates, defined as
\(R_{\text{e},\mathbf{s}}=\mathfrak r P_{\text{e},\mathbf{s}}\) and
\(R_{\text{a},\mathbf{s}}=\mathfrak r P_{\text{a},\mathbf{s}}\), respectively.
Here, \(\mathfrak r=\Delta N/\Delta t\) denotes the atomic injection rate, with
\(\Delta N\) representing the number of atoms injected during the interval
\(\Delta t\). (See Ref.~\cite{Ordonez2025QAspST} for details.) 

We now focus on the evaluation of the emission rate near the outer horizon by
analyzing the integral~\eqref{probem1}, which defines the quantum amplitude for
this process, with the outgoing \((+)\) CQM mode of
Eq.~\eqref{eq:CQM_modes}. The emission amplitude becomes
\begin{equation}
\setlength\abovedisplayskip{13pt}
\begin{aligned}
\int d\tau\, [\phi^{+(\mathrm{CQM})}_{\mathbf{s}}(t(\tau), \mathbf{r}(\tau))]^*
e^{i\nu\tau} 
& = 
 \int d\tau\, x^{-i\Theta} e^{-im\tilde{\varphi}}
 e^{i\tilde{\omega}t} e^{i\nu\tau}
  \\
&  \overset{(\mathcal{H})}{\sim}
 k \int_0^{x_\circ} dx\, x^{-i\tilde{\omega}/\kappa} e^{-iqx}.
\label{eqintm1}
\end{aligned}
\setlength\belowdisplayskip{13pt}
\end{equation}
where \(q=C\tilde{\omega}+k\nu+\alpha m\), and \(x_\circ\) denotes the radial
cutoff delimiting the near-horizon region.

The behavior of the integral in Eq.~\eqref{eqintm1} is governed by the
interplay between the logarithmic phase
\(x^{-i\tilde{\omega}/\kappa}\) and the oscillatory factor
\(e^{-iqx}\). The logarithmic phase arises from the conformal structure of
the near-horizon region and reflects the scale invariance of the CQM system,
characterized by the dimensionless parameter
\(\Theta=\tilde{\omega}/(2\kappa)\). Following the Kerr analysis of
Ref.~\cite{Ordonez2025QAspST}, we work in the large-atomic-frequency regime
\(\nu\gg\tilde{\omega}\) and \(\nu\gg|\alpha m|\). In this regime, the
\(k\nu\) term dominates \(q\), so that \(q\simeq k\nu\).

Away from the horizon, the factor \(e^{-iqx}\) is highly oscillatory, while
the logarithmic factor changes only weakly over successive oscillations. The
contributions from this part of the integration region therefore average out
and do not determine the leading result. Near the horizon, by contrast, the
scale-invariant logarithmic phase controls the leading nonzero contribution.
Consequently, the upper limit \(x_\circ\) in Eq.~\eqref{eqintm1} can be
extended to infinity at leading order. The resulting oscillatory integral is defined by introducing a small positive
convergence parameter \(\epsilon>0\) through
\(\nu\to\nu-i\epsilon\), with the limit \(\epsilon\to0^+\) taken only after
the integration has been performed. We then obtain
\begin{equation}
\begin{aligned}
\int d\tau\,
[\phi^{+(\mathrm{CQM})}_{\mathbf{s}}
(t(\tau),\mathbf r(\tau))]^*
e^{i\nu\tau}
&\simeq
k\lim_{\epsilon\to0^+}
\int_0^\infty dx\,
x^{-i\tilde{\omega}/\kappa}
e^{-iqx-k\epsilon x}
\\
&=
k(iq)^{-1+i\tilde{\omega}/\kappa}
\Gamma\!\left(
1-\frac{i\tilde{\omega}}{\kappa}
\right).
\end{aligned}
\end{equation}
The convergence parameter serves only to define the oscillatory integral and
has no additional physical significance.

Taking the modulus squared and using \(q\simeq k\nu\) in the
large-atomic-frequency regime specified above, the emission rate becomes
\begin{equation}
R_{\text{e},\mathbf{s}} 
= \frac{2\pi \mathfrak{r} g^2 \tilde{\omega}}{\kappa \nu^2} 
\left( \frac{1}{e^{2\pi\tilde{\omega}/\kappa} - 1} \right).
\label{emrate}
\end{equation}
Within the same approximation, the corresponding absorption rate
\(R_{\text{a},\mathbf{s}}\) can be obtained analogously or directly from
Eq.~\eqref{emrate} by applying the substitution
\(\tilde{\omega}\to-\tilde{\omega}\), yielding
\begin{equation}
R_{\text{a},\mathbf{s}} 
= \frac{2\pi \mathfrak{r} g^2 \tilde{\omega}}{\kappa \nu^2} 
\left( \frac{1}{1 - e^{-2\pi \tilde{\omega}/\kappa}} \right)
= e^{2\pi\tilde{\omega}/\kappa} R_{\text{e},\mathbf{s}}.
\label{remrabs12}
\end{equation}
For the ingoing \((-)\) CQM modes, the logarithmic phases cancel, so these
modes do not contribute to the leading near-horizon emission and absorption
rates~\cite{camblong2020near}. Therefore, unless otherwise stated, the emission
and absorption rates throughout this work refer exclusively to the outgoing
\((+)\) CQM modes.

This analysis of the transition amplitudes shows that the logarithmic nature
of the CQM waves in Eq.~\eqref{eq:CQM_modes}, which display a Russian-doll
symmetry~\cite{camblong2005black,camblong2020near}, governs the universal
near-horizon structure of the emission and absorption rates. As a result of
this conformal dominance, the thermal nature of the radiation field follows,
as discussed next.

\subsection{Thermal radiation-field dynamics and steady state}

The emission rate of scalar quanta by the atoms, given in Eq.~\eqref{emrate},
exhibits a Planckian form, suggesting an underlying thermal nature. In fact,
the ratio between the emission and absorption rates yields
\begin{equation}
    \frac{R_{\text{e},\mathbf{s}}}{R_{\text{a},\mathbf{s}}} = e^{-2\pi\tilde{\omega}/\kappa},
    \label{balance11}
\end{equation}
which matches the detailed balance condition for a thermal state governed by
the Boltzmann factor,
\begin{equation}
    \frac{R_{\text{e},\mathbf{s}}}{R_{\text{a},\mathbf{s}}} = e^{-\beta\tilde{\omega}},
\end{equation}
with an effective temperature
\begin{equation}
    T = \beta^{-1} = \frac{\kappa}{2\pi} = \beta_H^{-1} = T_H.
    \label{temp12}
\end{equation}
Thus, the effective temperature coincides precisely with the Hawking
temperature of the black hole.

To make this thermal statement explicit at the level of the radiation-field
density matrix, we consider the composite system formed by the atomic cloud and
the radiation field of scalar quanta. For this composite system, with density
matrix \(\hat{\rho}^{\rm rad\text{-}atom}\), the corresponding reduced density
matrix of the radiation-field subsystem is obtained by partial tracing over
the atomic degrees of freedom:
\begin{equation}
    \hat{\rho}^{\rm rad}=\Tr_{\rm atom}\!\left[\hat{\rho}^{\rm rad\text{-}atom}\right].
\end{equation}
For the remainder of the paper, when referring to density matrices, we omit
the superscript `rad', with the understanding that the analysis is focused
only on the radiation field.

A key property of this setup is that, for a cloud of freely falling atoms
injected at random times, the radiation-field density matrix becomes diagonal
in the occupation number basis. This can be shown by an appropriate averaging
procedure with respect to an injection distribution function. The result is a
multimode master equation 
\begin{eqnarray}
\dot{\rho}_{\text{diag}}\!\qty(\!\{n\}\!) 
&=& - \sum_j \!\Big[
R_{\text{e},j} \qty[(n_j+1)\rho_{\text{diag}}(\{n\}) - n_j \rho_{\text{diag}}(\{n\}_{n_j-1})] \nonumber\\
&&\hspace{1.5cm}+ R_{\text{a},j} \qty[n_j \rho_{\text{diag}}(\{n\}) - (n_j+1) \rho_{\text{diag}}(\{n\}_{n_j+1})]
\Big],
\label{fmeq12}
\end{eqnarray}
whose structure is general and independent of the specific background geometry.
(See Ref.~\cite{Ordonez2025QAspST}.) In Eq.~\eqref{fmeq12}, the index \(j\)
labels the field modes \(\mathbf{s}_j\), and the diagonal density matrix,
written in the occupation number basis
\(\{n\}\equiv\qty{n_1,n_2,\ldots,n_j,\ldots}\), is given by
\(\rho_{\text{diag}}(\{n\})=\rho_{n_1,n_2,\ldots;\,n_1,n_2,\ldots}\), with
\(n_j\equiv n_{\mathbf{s}_j}\) the occupation number of quanta in mode
\(\mathbf{s}_j\). The notation
\(\{n\}_{n_j+q}\equiv\qty{n_1,n_2,\ldots,n_j+q,\ldots}\) represents a shift
of \(q\in\mathbb Z\) in the \(j\)th mode occupation number, with all other
occupation numbers unchanged. The same master-equation structure can also be
used in a variety of physical problems that can be approximately
modeled~\cite{BEC-laser-analog_Scully-etal-2022,Ordonez2025QAspST} by
Eq.~\eqref{fmeq12}.

The steady-state solution \(\rho_{\text{diag}}^{(\text{SS})}(\{n\})\) is
obtained by setting the time derivative in Eq.~\eqref{fmeq12} to zero. For a
single mode, the steady-state distribution satisfies~\cite{Ordonez2025QAspST}
\begin{eqnarray}
    \left. \rho^{(\text{SS})}_{n_j,n_j}\right|_{\text{single-mode}}
    = \qty[1-\qty(\frac{R_{\text{e},j}}{R_{\text{a},j}})]
      \qty(\frac{R_{\text{e},j}}{R_{\text{a},j}})^{n_j}
    = \frac{1}{Z_j} e^{-n_j\beta\tilde{\omega}_j},
    \label{rhosss}
\end{eqnarray}
where \(Z_j=[1-e^{-\beta\tilde{\omega}_j}]^{-1}\) is the partition function
for mode \(\mathbf{s}_j\), and \(\beta=2\pi/\kappa\).

Because the atoms are injected randomly and the field modes evolve
independently, the full multimode steady-state density matrix factorizes as a
product of single-mode contributions:
\begin{eqnarray}
    \rho^{(\text{SS})}_{\text{diag}}(\{n\})
    = \prod_j \rho^{(\text{SS})}_{n_j,n_j},
\end{eqnarray}
from which the explicit thermal form follows:
\begin{eqnarray}
    \rho^{(\text{SS})}_{\text{diag}}(\{n\})
    = \mathcal N \prod_j \qty(\frac{R_{\text{e},j}}{R_{\text{a},j}})^{n_j}
    = \frac{1}{Z(\beta)} \prod_j e^{-n_j \beta \tilde{\omega}_j},
    \label{hbarre12}
\end{eqnarray}
where \(Z(\beta)=\mathcal N^{-1}=\prod_j Z_j
=\prod_j[1-e^{-\beta\tilde{\omega}_j}]^{-1}\) is the full partition function.

Therefore, the steady-state density matrix of the radiation field corresponds
to a thermal distribution at the Hawking temperature \(T_H=\kappa/(2\pi)\),
consistent with Eqs.~\eqref{balance11}--\eqref{temp12}. This density matrix
reproduces the expected Planckian distribution for the steady-state average
occupation number,
\begin{equation}
    \expval{n_j}^{(\text{SS})} = \frac{1}{e^{\beta\tilde{\omega}_j} - 1}.
\end{equation}
Crucially, this thermal behavior emerges directly from the near-horizon CQM
structure, which governs the emission and absorption processes.

\section{HBAR thermodynamics}
\label{sec:HBAR-thermo}

This section develops the thermodynamic content of the radiation-field density
matrix obtained above. In the near-steady regime, we examine the entropy and
corotating-energy relations associated with the HBAR process and their
connection with Kerr horizon thermodynamics.

\subsection{From the von Neumann formula to thermodynamic entropy}

The HBAR von~Neumann entropy~\cite{scully2018quantum} is obtained from the
standard entropy expression \(S_{\rm rad}=-\Tr[\hat{\rho}\ln\hat{\rho}]\). For
trace-preserving evolution, its time derivative defines the corresponding HBAR
entropy flux,
\(\dot S_{\rm rad}=-\Tr[\dot{\hat\rho}\ln\hat\rho]\). In the
occupation-number basis introduced above, this becomes
\begin{equation}
    \dot{S}_{\rm rad}
    =
    -\sum_{\{n\}}
    \dot{\rho}_{\text{diag}}(\{n\})
    \ln\!\qty[\rho_{\text{diag}}(\{n\})].
    \label{vnebar}
\end{equation}

Near the steady-state configuration, the logarithm in Eq.~\eqref{vnebar} can
be evaluated to leading order using the steady-state solution
\(\rho^{(\text{SS})}_{\text{diag}}(\{n\})\) given in
Eq.~\eqref{hbarre12}. In this regime, the entropy flux becomes
\begin{eqnarray}
    \dot{S}_{\rm rad} 
    &\simeq& -\sum_{\{n\}} \dot{\rho}_{\text{diag}}(\{n\})
    \ln\!\qty[\rho^{(\text{SS})}_{\text{diag}}(\{n\})] \nonumber\\
    &=& -\sum_{\{n\}} \dot{\rho}_{\text{diag}}(\{n\})
    \ln\!\qty[\frac{1}{Z(\beta)} \prod_j e^{-n_j \beta \tilde{\omega}_j}] \nonumber\\
    &=& \sum_j \sum_{\{n\}} \dot{\rho}_{\text{diag}}(\{n\})\, n_j \beta \tilde{\omega}_j 
    - \sum_j \sum_{\{n\}} \dot{\rho}_{\text{diag}}(\{n\})
    \ln(1 - e^{-\beta\tilde{\omega}_j}),
    \label{eqdhj1}
\end{eqnarray}
where \(Z(\beta)=\prod_j [1 - e^{-\beta \tilde{\omega}_j}]^{-1}\). 

Using the trace normalization
\(\Tr[\hat{\rho}]=\sum_{\{n\}}\rho_{\text{diag}}(\{n\})=1\) and the
definition of the instantaneous expectation value
\(\expval{n_j}=\sum_{\{n\}} n_j \rho_{\text{diag}}(\{n\})\),
Eq.~\eqref{eqdhj1} simplifies to
\begin{eqnarray}
    \dot{S}_{\rm rad} 
    &\simeq& \sum_j 
    \underbrace{\qty[\sum_{\{n\}} \dot{\rho}_{\text{diag}}(\{n\})\, n_j]}_{\dot{\expval{n_j}}} 
    \beta \tilde{\omega}_j 
    - \sum_j 
    \underbrace{\qty[\sum_{\{n\}} \dot{\rho}_{\text{diag}}(\{n\})]}_{=\,0}
    \ln(1 - e^{-\beta \tilde{\omega}_j}) \nonumber\\
    &=& \beta_H \sum_j \dot{\expval{n_j}}\, \tilde{\omega}_j.
    \label{eqdhj2}
\end{eqnarray}
Here the second term vanishes by trace conservation, while
\(\beta=\beta_H\) follows from Eq.~\eqref{temp12}.

The quantity \(\dot{\expval{n_j}}\,\tilde{\omega}_j\) represents the
contribution of mode \(j\) to the energy flux carried by the acceleration
radiation, measured with respect to the horizon-generating Killing field
\(\xi=\partial_t+\Omega_H\partial_\varphi\). Equivalently, it is the energy
flux measured in the frame corotating with the black hole. The total
corotating energy flux is therefore
\begin{eqnarray}
    \dot{\tilde{E}}_{\rm rad} 
    &=& \sum_j \dot{\expval{n_j}}\, \tilde{\omega}_j 
    =\sum_j \dot{\expval{n_j}}(\omega_j - \Omega_H m_j) \nonumber\\
    &=& \dot{E}_{\rm rad} - \Omega_H \dot{J}_{{\rm rad},z},
    \label{efluxti1}
\end{eqnarray}
where \(\dot{E}_{\rm rad}\) and \(\dot{J}_{{\rm rad},z}\) denote the energy and
axial angular momentum fluxes carried by the radiation field, respectively.

Therefore, in the same near-steady regime, the HBAR entropy flux
satisfies~\cite{Ordonez2025QAspST}
\begin{equation}
    \dot{S}_{\rm rad} 
    \simeq \beta_H (\dot{E}_{\rm rad} - \Omega_H \dot{J}_{{\rm rad},z})
    = \beta_H \dot{\tilde{E}}_{\rm rad}.
    \label{sdotedot1}
\end{equation}
At the level of variations, the same relation gives
\begin{equation}
    \delta S_{\rm rad} 
    \simeq \beta_H (\delta E_{\rm rad} - \Omega_H \delta J_{{\rm rad},z})
    \equiv \delta S^{(\text{th})}_{\rm rad}.
    \label{enth12}
\end{equation}
Here, \(\delta S^{(\mathrm{th})}_{\rm rad}\) denotes the thermodynamic entropy
variation of the radiation field, which coincides to leading order with the
variation of the HBAR entropy in the near-equilibrium regime.

Finally, the HBAR entropy itself can be evaluated within the same near-steady
approximation. To leading order, the logarithmic term can again be evaluated
using \(\rho^{(\text{SS})}_{\text{diag}}(\{n\})\). The same normalization and
expectation-value definitions then give
\begin{eqnarray}
    S_{\rm rad} 
    &\simeq& \sum_j \!\left[\sum_{\{n\}} \rho_{\text{diag}}(\{n\})\, n_j\right]\! \beta \tilde{\omega}_j 
    - \sum_j \!\left[\sum_{\{n\}} \rho_{\text{diag}}(\{n\})\right] 
    \ln(1 - e^{-\beta\tilde{\omega}_j}) \nonumber\\
    &=& \sum_j \langle n_j \rangle\, \beta \tilde{\omega}_j 
    - \sum_j \ln(1 - e^{-\beta \tilde{\omega}_j}) \nonumber\\
    &=& \beta_H (\tilde{E}_{\rm rad} - \tilde{F}_{\rm rad}) 
    \equiv S^{(\text{th})}_{\rm rad},
    \label{entrphif}
\end{eqnarray}
where \(\tilde{E}_{\rm rad}=\sum_j\langle n_j\rangle\,\tilde{\omega}_j\),
\(\beta=\beta_H\) in the thermal regime considered here, and
\(\tilde{F}_{\rm rad}\) is the Helmholtz free energy, with
\begin{equation}
\beta \tilde{F}_{\rm rad}
=
\sum_j \ln(1 - e^{-\beta \tilde{\omega}_j})
=
-\ln Z(\beta).
\end{equation}
Hence, the HBAR entropy obtained in Eq.~\eqref{entrphif} is consistent with the
flux expression in Eq.~\eqref{sdotedot1}, as its variation at fixed horizon
data \((\beta_H,\Omega_H)\) reduces to Eq.~\eqref{enth12}. We conclude that the
Kerr black hole behaves effectively as a thermal reservoir at the
Hawking temperature \(T_H\), a property that emerges naturally from the
quantum-optical treatment through the emission and absorption ratios encoded
in the master equation~\eqref{fmeq12}.

\subsection{HBAR-black-hole thermodynamic correspondence}

The thermodynamic variation of the Bekenstein--Hawking entropy of a rotating
black hole in terms of its mass \(M\) and angular momentum \(J\) is given
by~\cite{bardeen1973four,frolov2012black}
\begin{equation}
    \delta S_{\text{BH}}
    =
    \beta_H
    \left(
        \delta M-\Omega_H\delta J
    \right).
\end{equation}
This relation has the same thermodynamic structure as the near-steady HBAR
entropy-flux and entropy-variation relations in Eqs.~\eqref{sdotedot1} and
\eqref{enth12}, respectively. Moreover, Eq.~\eqref{temp12} identifies the
inverse temperature of the radiation field with that of the black hole,
\(\beta=\beta_H\). This structural agreement therefore establishes the formal
thermodynamic correspondence~\cite{Ordonez2025QAspST}
\begin{equation}
    (S_{\rm rad}, E_{\rm rad}, J_{{\rm rad},z})
    \;\overset{\beta=\beta_H}{\longleftrightarrow}\;
    (S_{\text{BH}}, M, J).
    \label{thercrs1}
\end{equation}

\subsection{Area-entropy-flux relation and radiation correspondence}
\label{sec:entropy-area-relation}

The Bekenstein--Hawking entropy-area relation is
given by~\cite{bekenstein1973black,hawking1975particle}
\begin{equation}
    \delta S_{\text{BH}} = \frac{1}{4}\, \delta A .
    \label{bhenalaw}
\end{equation}
Here \(A\) denotes the dimensionless horizon area, namely the geometric area
measured in Planck-area units. The first law of black-hole mechanics then
gives~\cite{bardeen1973four}
\begin{equation}
    \delta A = 4 \beta_H (\delta M - \Omega_H \delta J).
    \label{deltaAdeltaM}
\end{equation}

Combining the HBAR-black-hole thermodynamic correspondence in
Eq.~\eqref{thercrs1} with the Bekenstein--Hawking area
law~\eqref{bhenalaw} yields, in the same near-steady regime, the analogous
entropy-area-flux relation
\begin{equation}
    \dot{S}_{\rm rad} \simeq \frac{1}{4}\, |\dot{A}_{\rm rad}|,
    \label{abssdot}
\end{equation}
where \(|\dot{A}_{\rm rad}|\) denotes the magnitude of the radiative
contribution to the horizon-area rate associated with the emitted scalar
quanta. The absolute value reflects the fact that this contribution decreases
the horizon area, whereas the corresponding radiation entropy flux is
positive. This does not conflict with the generalized second law of
thermodynamics, which constrains the sum of the black-hole entropy and the
entropy outside the horizon, including the atomic and radiative
sectors~\cite{bekenstein1974generalized}; see
Ref.~\cite{Ordonez2025QAspST} for a detailed discussion in the HBAR
framework.

Applying Eq.~\eqref{deltaAdeltaM} to the radiative contribution and using the
HBAR-black-hole thermodynamic correspondence~\eqref{thercrs1} yields the
power-area relation~\cite{Ordonez2025QAspST}
\begin{equation}
    |\dot{A}_{\rm rad}| = 4 \beta_H \dot{\tilde{E}}_{\rm rad}.
    \label{a4bete1}
\end{equation}

The near-steady area-entropy-flux statement~\eqref{abssdot}, together with its
thermodynamic relation to the corotating energy flux,
Eqs.~\eqref{deltaAdeltaM} and \eqref{a4bete1}, provides the bridge to the
information-theoretic concepts developed below. In the remaining sections,
these connections will be formulated as a geometric information-theoretic
framework for the HBAR channel and expressed in physical units by restoring
the constants \(c\), \(\hbar\), \(k_B\), and \(G\).

\section{Entropy-area law from data processing (Spohn's theorem perspective)}
\label{sec:entropy-area-Spohn}

In the previous section, we showed that the Kerr black hole effectively behaves
as a thermal reservoir. We now reexamine this thermodynamic result from the
perspective of the dynamical evolution of the reduced density matrix
\(\hat{\rho}\) of the radiation field, while maintaining the weak atom--field
interaction assumption introduced in Sec.~\ref{sec:HBAR-Kerr}. This
perspective is naturally formulated within irreversible thermodynamics for
quantum systems weakly coupled to reservoirs, following
Refs.~\cite{Spohn1,Spohn2}. As before, we omit the superscript `rad' on the
reduced density matrix, with the understanding that the analysis is focused
only on the radiation field. Within the Markovian weak-coupling description
considered here, the reduced density matrix evolves according to
\(\dot{\hat{\rho}}(t)=\mathcal{L}[\hat{\rho}(t)]\), where
\(\mathcal{L}\) is the generator of the evolution. We denote by
\(\hat{\rho}^{(\mathrm{SS})}\) the faithful stationary density matrix satisfying
\(\mathcal{L}[\hat{\rho}^{(\mathrm{SS})}]=0\). The entropy production rate can
then be introduced as follows.

The entropy balance equation in irreversible thermodynamics reads
\begin{equation}
\frac{d S}{d t} = - {I}_S
+ \sigma ,
\end{equation}
where \(S\) is the entropy, \({I}_S\) is the net outward entropy current, and
\(\sigma\) is the entropy production rate. This is the volume-integrated form
of the local continuity equation
\(\partial s/\partial t = - \mathrm{div}\,\mathbf{J}_S + \sigma'\), where
\(s\), \(\mathbf{J}_S\), and \(\sigma'\) are the entropy density,
entropy-current density, and entropy-production density,
respectively~\cite{degroot1984nonequilibrium}. In the integrated form,
\(S=\int_V s\,dV\),
\({I}_S=\int_{\partial V}\mathbf J_S\cdot d\mathbf A\), and
\(\sigma=\int_V\sigma'\,dV\).

For the quantum Markovian evolution considered here, the entropy production
rate can be written in terms of the relative entropy with respect to the
steady-state density matrix:
\begin{equation}
\sigma(\hat{\rho}) =
-\frac{d}{dt}\,\mathrm{Tr}\!\left[\hat{\rho}(t)
\big(\ln\hat{\rho}(t)-\ln\hat{\rho}^{(\mathrm{SS})}\big)\right]
=
-\frac{d}{dt}\,
S\!\left(\hat{\rho}(t)\,\big\|\,\hat{\rho}^{(\mathrm{SS})}\right),
\end{equation}
where \(S(\cdot\|\cdot)\) denotes the quantum relative
entropy~\cite{lindblad1973entropy}.

According to Spohn's theorem~\cite{Spohn1,Spohn2}, the entropy production is
nonnegative, \(\sigma(\hat{\rho})\ge0\), for a quantum Markov semigroup with a
faithful stationary density matrix. The Markovian assumption is essential to
this monotonicity statement; entropy production under more general
non-Markovian dynamical maps requires a separate
analysis~\cite{Marcantonietal2017}. In the present weak-coupling setting, the
random injection of atoms renders the reduced density matrix diagonal in the
occupation-number basis and leads to the Markovian master
equation~\eqref{fmeq12}, whose faithful thermal stationary solution is given in
Eq.~\eqref{hbarre12}. The nonnegativity of \(\sigma(\hat{\rho})\) is the
infinitesimal form of the data-processing inequality for quantum relative
entropy~\cite{lindblad1975completely,ahlswede2001quantum}. Spohn's inequality
therefore becomes
\begin{equation}
\sigma(\hat{\rho}) =
-\sum_{\{n\}} \dot{\rho}_{\mathrm{diag}}(\{n\})\,
\Big[
\ln\rho_{\mathrm{diag}}(\{n\})
-
\ln\rho^{(\mathrm{SS})}_{\mathrm{diag}}(\{n\})
\Big]
\ge 0.
\label{spohn-ineq}
\end{equation}

Using the thermal steady-state density matrix at inverse temperature
\(\beta_H\),
\begin{equation}
\rho^{(\mathrm{SS})}_{\mathrm{diag}}(\{n\})
=
\frac{1}{Z(\beta_H)}
\prod_j e^{-n_j\beta_H\tilde{\omega}_j},
\end{equation}
together with Eqs.~\eqref{vnebar}--\eqref{efluxti1},
Eq.~\eqref{spohn-ineq} gives
\begin{equation}
\sigma(t) =
\dot{S}_{\rm rad}
-
\beta_H\,\dot{\tilde{E}}_{\rm rad}
\ge 0,
\label{second-law}
\end{equation}
where \(\dot{\tilde{E}}_{\rm rad}\) denotes the energy flux measured in the
frame corotating with the horizon. Thus, Eq.~\eqref{second-law} generalizes the
near-steady relation~\eqref{sdotedot1} by retaining the additional nonnegative
entropy-production term \(\sigma(t)\) away from the thermally saturated
regime.

In the near-steady, thermally saturated regime, the entropy production
contribution vanishes to leading order, and Eq.~\eqref{second-law} reduces to
\begin{equation}
\dot{S}_{\rm rad}
\simeq
-\sum_{\{n\}} \dot{\rho}_{\mathrm{diag}}(\{n\})
\ln\!\left[
\rho^{(\mathrm{SS})}_{\mathrm{diag}}(\{n\})
\right]
=
\beta_H\,\dot{\tilde{E}}_{\rm rad}.
\label{Spohn-equality}
\end{equation}
Using the relation between the energy flux and the horizon-area rate associated
with the emitted scalar quanta, Eq.~\eqref{a4bete1}, we rederive the
entropy-area equation~\eqref{abssdot}, now as a particular case.

At this point, it is useful to work in general physical units by restoring the
fundamental constants \(\hbar\), \(c\), \(G\), and \(k_B\). Accordingly, in the
remainder of the paper we distinguish the dimensionful physical entropy
\(S_{\rm phys}\) from the dimensionless information-theoretic entropy
\(S=S_{\rm phys}/k_B\). For the HBAR entropy considered here, we write
\(S\equiv S^{(\rm rad)}\), so that
\(S_{\rm phys}^{(\rm rad)}=k_B S^{(\rm rad)}\). The near-steady entropy-flux
relation then becomes
\begin{equation}
\dot{S}^{(\rm rad)}
\simeq
\frac{1}{k_B T_H}\,\dot{\tilde{E}}_{\rm rad}
=
\frac{1}{4 \ell_P^2}\,\big|\dot{A}_{\rm rad}\big|,
\label{Srate_landauer}
\end{equation}
where
\begin{equation}
\ell_P= \sqrt{\frac{\hbar G}{c^3}},
\label{eq:Planck-length}
\end{equation}
so that \(\ell_P^2\) gives the elementary Planck area. In
Eq.~\eqref{Srate_landauer}, \(|\dot A_{\rm rad}|\) denotes the magnitude of
the geometric horizon-area rate in physical units. Equivalently,
\(|\dot A_{\rm rad}|/\ell_P^2\) is the corresponding dimensionless area rate
measured in Planck-area units.

The relation in Eq.~\eqref{Srate_landauer} should be understood as the
near-steady, thermally saturated limit of the Spohn entropy-production
balance. Away from this regime, Spohn's theorem gives the nonnegative
entropy-production term in Eq.~\eqref{second-law}, so that the entropy balance
is modified by additional irreversible production. In what follows,
\(\simeq\) denotes equality at this leading order, while \(\lesssim\) and
\(\gtrsim\) denote the corresponding upper and lower bounds.

In this way, Eq.~\eqref{second-law} expresses the second law for the HBAR
channel as the infinitesimal form of the data-processing inequality for quantum
relative entropy, thereby establishing the nonnegativity of entropy production.
Its near-steady-state saturation, Eq.~\eqref{Spohn-equality}, identifies the
radiation flux as thermodynamically consistent with a reservoir at the Hawking
temperature. In this saturated regime, Eq.~\eqref{Srate_landauer} provides the
geometric realization of the same thermodynamic balance: the HBAR entropy flux
is directly proportional to the magnitude of the radiative contribution to the
horizon-area rate, with the characteristic factor \(1/4\) of the
Bekenstein--Hawking entropy-area law. Together, these relations establish the
internal thermodynamic consistency of the HBAR process and provide a geometric
interpretation of its near-steady entropy balance.
\section{Area-cost law for communication}
\label{sec:bits_area}

In this section, we first review the standard Bekenstein--Holevo route to an
information-rate bound. We then combine the Holevo entropy bound with the HBAR
entropy-area relation to obtain an area-cost law for classical communication,
together with its Landauer-type energetic interpretation.

\subsection{Bekenstein bound for communication rate}

Let us consider a finite physical system of total energy \(E_\circ\) confined
within a sphere of radius \(R_\circ\). Its thermodynamic entropy is limited by
the available energy through the Bekenstein
bound~\cite{Bekenstein1,bekenstein1981universal,bekenstein1981energy},
\begin{equation}
S_{\mathrm{phys}} \le \frac{2\pi k_B R_\circ E_\circ}{\hbar c} .
\label{BekensteinBound}
\end{equation}
Here \(S_{\mathrm{phys}}\) denotes the physical entropy, and the conversion to
bits will be implemented through the usual factor \(1/\ln(2)\).

The accessible classical information \(I_{\mathrm{class}}\), measured in bits,
encoded in a physical ensemble is constrained by the entropy through the
Holevo bound~\cite{Holevo1973,NielsenChuang2010,Wilde2017},
\begin{equation}
I_{\mathrm{class}} \le \frac{S_{\mathrm{phys}}}{k_B \ln(2)} .
\label{holevo12}
\end{equation}

For a message of size \(I_{\mathrm{class}}\) transmitted during a time interval
\(\tau\), the magnitude of the average information rate
satisfies~\cite{bekenstein1981energy}
\begin{equation}
\big|\dot{I}_{\mathrm{class}}\big|
\equiv
\frac{I_{\mathrm{class}}}{\tau}
\le
\frac{1}{k_B \ln(2)}\,\frac{S_{\mathrm{phys}}}{\tau}
\le
\frac{1}{k_B \ln(2)}\,\frac{S_{\mathrm{phys}}^{\max}}{\tau},
\label{IrateStart}
\end{equation}
where \(S_{\mathrm{phys}}^{\max}\) is bounded by
Eq.~\eqref{BekensteinBound}. Causality further requires that the signaling time
cannot be shorter than the light-crossing time,
\(\tau \ge 2R_\circ/c\). Substituting this lower bound on \(\tau\) into
Eq.~\eqref{IrateStart} gives
\begin{equation}
\big|\dot{I}_{\mathrm{class}}\big|
\le
\frac{\pi E_\circ}{\hbar \ln(2)},
\label{Bremermann}
\end{equation}
which constrains the maximal information-processing rate of any physical
system by its available energy.

\subsection{Area-cost law for communication in the HBAR channel}

We now apply this reasoning to the emission of acceleration radiation by the
atomic cloud near the black-hole horizon. From the near-steady regime analysis
and the saturated Spohn entropy balance in Eq.~\eqref{Srate_landauer}, the
physical entropy rate of the radiation field is, to leading order,
\begin{equation}
\dot{S}_{\rm phys}^{(\rm rad)}
\simeq
\frac{k_B}{4\ell_P^2}
\,\big|\dot{A}_{\rm rad}\big|.
\label{Srate_landauer_final}
\end{equation}

Consider a communication interval \([t_i,t_f]\), and let
\(I_{\mathrm{class}}\) denote the accessible classical information encoded in
the outgoing radiation produced during this interval. The Holevo bound applies
to the ensemble of radiation-field density matrices associated with the
complete finite process. Assuming that the outgoing radiation initially
carries negligible entropy and no encoded classical information, we obtain
\begin{equation}
I_{\mathrm{class}}
\leq
\frac{S_{\rm phys}^{(\rm rad)}(t_f)}
{k_B\ln(2)}
\simeq
\frac{\Delta S_{\rm phys}^{(\rm rad)}}
{k_B\ln(2)}
\simeq
\frac{1}{4\ln(2)\,\ell_P^2}
\int_{t_i}^{t_f}
\big|\dot A_{\rm rad}(t)\big|\,dt.
\label{bits_per_area_rate_final}
\end{equation}
The last relation uses the near-steady, thermally saturated entropy-area
balance over the same emission interval. Thus, the accessible classical
information supported by the emitted radiation is bounded, at the order
considered, by the accumulated radiative horizon-area budget.

Assuming that the area change has a definite sign over the communication
interval, Eq.~\eqref{bits_per_area_rate_final} gives
\begin{equation}
I_{\mathrm{class}}
\lesssim
\frac{1}{4\ln(2)\,\ell_P^2}
\,\big|\delta A_{\rm rad}\big|.
\label{area_budget_bits}
\end{equation}
Accordingly, the area cost per accessible classical bit obeys the
leading-order lower bound
\begin{equation}
\big|\delta A_{\rm rad}\big|_{\mathrm{per\,bit}}
\gtrsim
4\ln(2)\,\ell_P^2.
\label{area_per_bit_final}
\end{equation}
Using
\(\ell_P^2=\hbar G/c^3\approx2.61\times10^{-70}\,\mathrm{m}^2\),
the lower-bound area scale in Eq.~\eqref{area_per_bit_final} is
\begin{equation}
4\ln(2)\,\ell_P^2
\approx
7.24\times10^{-70}\ \mathrm{m}^2.
\label{area_per_bit_numeric}
\end{equation}

Equation~\eqref{area_per_bit_final} expresses a Landauer-type principle for
the geometric cost of communication: any accessible classical bit carried by
the idealized HBAR channel requires, at this order, a horizon-area reduction
of at least \(4\ln(2)\,\ell_P^2\). This lower bound is independent of the
black-hole mass and of other microscopic details.
\subsection{Energetic interpretation: Landauer-type limit}

The corresponding Landauer-type lower bound on the corotating energy per
accessible bit follows by combining the Holevo bound with the near-steady
thermodynamic relation in Eq.~\eqref{Srate_landauer}:
\begin{equation}
\left.\delta\tilde{E}_{\rm rad}\right|_{\mathrm{per\,bit}}
\gtrsim
k_B T_H\ln(2).
\label{Landauer_energy_bit}
\end{equation}
The coefficient \(k_B T_H\ln(2)\) gives the ideal leading-order minimum
obtained by saturating the Holevo bound within the near-steady thermodynamic
relation. This
result parallels Landauer's principle for information erasure,
\(\delta E_{\min}=k_B T\ln(2)\). Here, however, the cost does not correspond
to erasing a bit, but rather to the minimum corotating-energy budget required
to support one accessible bit in the HBAR channel.

For a nonextremal Kerr black hole, the Hawking temperature depends on both its
mass and rotation and vanishes in the extremal limit. Accordingly, the
energetic lower bound per transmitted bit is smaller for lower-temperature
nonextremal Kerr backgrounds.

The HBAR process therefore reveals a dual relation between geometry and
information: the geometric lower bound per accessible bit,
Eq.~\eqref{area_per_bit_final}, is fixed in Planck-area units, while the
corresponding energetic lower bound, Eq.~\eqref{Landauer_energy_bit}, is
governed by the Hawking temperature. Together, these relations express the
thermodynamic-informational consistency of the horizon as a communication
channel, where the emission of scalar quanta carries both an energetic and a
geometric cost for information transfer.
\section{Area-cost law for correlations: mutual-information perspective}
\label{sec:mi_area}

Let \(\mathcal{R}\) denote the outgoing acceleration radiation, and let
\(\mathcal{E}\) denote everything else: the black hole, the atoms, and the
ancillary environment. Mutual-information analysis provides a general measure
of the correlations shared by two subsystems~\cite{batina2011mutual}. Thus,
the total correlations generated between \(\mathcal{R}\) and
\(\mathcal{E}\) are quantified by the mutual information
\begin{equation}
I_M(\mathcal{R}\!:\!\mathcal{E})
=
S(\mathcal{R}) + S(\mathcal{E}) - S(\mathcal{R}\mathcal{E}),
\label{MI_def}
\end{equation}
where \(S(\cdot)\) is the von~Neumann entropy, measured as a dimensionless
quantity in nats.

The Araki--Lieb inequality~\cite{araki1970entropy}, together with
subadditivity, constrains the entropy of composite quantum systems according to
\begin{equation}
\big|S(\mathcal{R}) - S(\mathcal{E})\big|
\le
S(\mathcal{R}\mathcal{E})
\le
S(\mathcal{R}) + S(\mathcal{E}).
\end{equation}
Substituting the lower bound into Eq.~\eqref{MI_def} gives
\begin{align}
I_M(\mathcal{R}\!:\!\mathcal{E})
&=
S(\mathcal{R}) + S(\mathcal{E}) - S(\mathcal{R}\mathcal{E})
\nonumber\\
&\le
S(\mathcal{R}) + S(\mathcal{E})
-
\big|S(\mathcal{R})-S(\mathcal{E})\big|
=
2 \min\{S(\mathcal{R}),S(\mathcal{E})\}.
\end{align}
Since
\(2\min\{S(\mathcal{R}),S(\mathcal{E})\}\le 2S(\mathcal{R})\), we obtain
the useful bound
\begin{equation}
I_M(\mathcal{R}\!:\!\mathcal{E})
\le
2S(\mathcal{R}).
\label{I_le_2S}
\end{equation}

Physically, Eq.~\eqref{I_le_2S} states that the total correlations, classical
and quantum, between the radiation field \(\mathcal{R}\) and its complementary
environment \(\mathcal{E}\) cannot exceed twice the entropy of
\(\mathcal{R}\). The sharper Araki--Lieb form shows that the true upper bound
is twice the entropy of the smaller subsystem. This bound is saturated when
the global density matrix describes a pure bipartite state, with
\(S(\mathcal{R})=S(\mathcal{E})\), in which case
\(I_M(\mathcal{R}\!:\!\mathcal{E})=2S(\mathcal{R})=2S(\mathcal{E})\). The
result is purely information-theoretic and follows only from quantum entropy
inequalities, independently of the details of the HBAR dynamics.

\subsection{HBAR analysis}

In the HBAR framework, the dimensionless entropy rate of the radiation field,
\(\dot S^{(\rm rad)}\), is related to the horizon-area rate associated with the
emitted scalar quanta in the near-steady saturated regime, as given in
Eq.~\eqref{Srate_landauer}. Thus,
\begin{equation}
\dot S^{(\rm rad)}(t)
\simeq
\frac{1}{4\ell_P^2}
\big|\dot A_{\rm rad}(t)\big|.
\end{equation}
Integrating over the emission interval \([t_i,t_f]\) gives
\begin{equation}
S^{(\rm rad)}(t_f)-S^{(\rm rad)}(t_i)
\simeq
\frac{1}{4\ell_P^2}
\int_{t_i}^{t_f}
\big|\dot A_{\rm rad}(t)\big|\,dt .
\end{equation}
If the outgoing radiation field has negligible initial entropy, so that
\(S^{(\rm rad)}(t_i)\simeq 0\), then
\begin{equation}
S(\mathcal{R})
=
S^{(\rm rad)}(t_f)
\simeq
\frac{1}{4\ell_P^2}
\int_{t_i}^{t_f}
\big|\dot A_{\rm rad}(t)\big|\,dt .
\end{equation}
When the area change has a definite sign during the emission process,
\begin{equation}
\int_{t_i}^{t_f}
\big|\dot A_{\rm rad}(t)\big|\,dt
=
|\delta A_{\rm rad}|,
\end{equation}
and therefore
\begin{equation}
S(\mathcal{R})
\simeq
\frac{|\delta A_{\rm rad}|}{4\ell_P^2}.
\end{equation}

Combining this entropy-area relation with the mutual-information
bound~\eqref{I_le_2S} yields the correlation-area constraint
\begin{equation}
I_M(\mathcal{R}\!:\!\mathcal{E})
\lesssim
\frac{|\delta A_{\rm rad}|}{2\ell_P^2}.
\label{MI_area_diff}
\end{equation}
Thus, the total correlations generated between the radiation field
\(\mathcal{R}\) and its environment \(\mathcal{E}\) are bounded by the finite
geometric area budget associated with the emitted scalar quanta.

The factor of two in Eq.~\eqref{MI_area_diff} is inherited directly from the
general bound~\eqref{I_le_2S}. Under the pure-state saturation condition
discussed above, each unit of entropy gained by \(\mathcal{R}\) is mirrored in
\(\mathcal{E}\), so that the mutual information is twice the single-sided
entropy. Consequently, the lower-bound area scale per bit of mutual
information is one half of that obtained for an accessible classical bit.

Equation~\eqref{MI_area_diff} is expressed in nats. Converting to bits by
dividing by \(\ln(2)\) gives
\begin{equation}
I_M^{(\mathrm{bits})}(\mathcal{R}\!:\!\mathcal{E})
\lesssim
\frac{1}{2\ln(2)\,\ell_P^2}
\,|\delta A_{\rm rad}|.
\label{MI_bits_diff}
\end{equation}
Accordingly, the horizon-area cost associated with the creation of one bit of
total radiation--environment mutual information obeys the leading-order lower
bound
\begin{equation}
|\delta A_{\rm rad}|
\gtrsim
2\ln(2)\,\ell_P^2 .
\label{area_per_bit_MI}
\end{equation}
The corresponding lower-bound area scale is
\(2\ln(2)\ell_P^2\approx3.62\times10^{-70}\,\mathrm{m}^2\), which is one half
of the lower-bound scale per accessible classical bit in
Eq.~\eqref{area_per_bit_final}.

Finally, any local processing of the radiation field \(\mathcal{R}\) cannot
increase its mutual information with \(\mathcal{E}\), by the data-processing
inequality. Hence, Eq.~\eqref{MI_area_diff} also bounds the
radiation--environment correlations that remain after any such processing:
the area budget constrains both the classically accessible information and the
total correlations that can be shared.

\subsection{Energetic interpretation: finite-process correlation budget}

The preceding area bound can also be expressed as an energy budget. In the
near-steady HBAR regime, the physical-unit form of Eq.~\eqref{Srate_landauer}
gives the dimensionless entropy flux of the radiation field as
\begin{equation}
\dot S^{(\rm rad)}
\simeq
\frac{\dot{\tilde E}_{\rm rad}}{k_B T_H}.
\end{equation}
Integrating over the emission interval \([t_i,t_f]\), and assuming that
\(T_H\) is approximately constant over this interval, gives
\begin{equation}
S^{(\rm rad)}(t_f)-S^{(\rm rad)}(t_i)
\simeq
\frac{1}{k_B T_H}
\int_{t_i}^{t_f}
\dot{\tilde E}_{\rm rad}(t)\,dt .
\end{equation}
If the outgoing radiation field has negligible initial entropy and the
corotating energy carried by the emitted scalar quanta is positive over the
interval considered, then
\begin{equation}
S(\mathcal{R})
\simeq
\frac{\delta\tilde E_{\rm rad}}{k_B T_H},
\end{equation}
where
\begin{equation}
\delta\tilde E_{\rm rad}
=
\int_{t_i}^{t_f}
\dot{\tilde E}_{\rm rad}(t)\,dt .
\end{equation}

Combining this finite entropy-energy budget with the Araki--Lieb
mutual-information bound gives
\begin{equation}
I_M(\mathcal{R}\!:\!\mathcal{E})
\lesssim
\frac{2\delta\tilde E_{\rm rad}}{k_B T_H}.
\end{equation}
Converting to bits using
\(I_M^{(\mathrm{bits})}=I_M/\ln(2)\), we obtain
\begin{equation}
I_M^{(\mathrm{bits})}(\mathcal{R}\!:\!\mathcal{E})
\lesssim
\frac{2\delta\tilde E_{\rm rad}}
{\ln(2)\,k_B T_H}.
\end{equation}
Therefore, the energy budget required to support
\(I_M^{(\mathrm{bits})}(\mathcal{R}\!:\!\mathcal{E})\) bits of
radiation-environment mutual information satisfies
\begin{equation}
\delta\tilde E_{\rm rad}
\gtrsim
\frac{k_B T_H\ln(2)}{2}
I_M^{(\mathrm{bits})}(\mathcal{R}\!:\!\mathcal{E}).
\end{equation}
For one bit of total radiation-environment mutual information, the required
corotating energy therefore satisfies
\begin{equation}
\left.\delta\tilde E_{\rm rad}\right|_{\mathrm{per\,bit}}
\gtrsim
\frac{k_B T_H\ln(2)}{2}.
\end{equation}
The coefficient \(k_BT_H\ln(2)/2\) gives the ideal leading-order minimum
obtained by saturating the mutual-information bound within the near-steady
entropy-energy relation. The factor of one-half reflects the bipartite correlation
accounting
\(I_M(\mathcal{R}\!:\!\mathcal{E})\leq2S(\mathcal{R})\): in the ideal
pure-state case, one unit of radiation entropy can correspond to two units of
radiation-environment mutual information.

\subsection{Reliability-aware area law from Fano's inequality}
\label{sec:fano_area}

Consider communication over the HBAR channel using a message
\(\mathfrak{M}\) drawn from a finite set \(\mathcal{M}\), of size
\(|\mathcal{M}|\), according to a probability distribution, and decoded as
\(\hat{\mathfrak M}\) with average error probability \(P_{\mathrm{error}}\).
Fano's inequality~\cite{Fano1961,HanVerdu1994} provides a lower bound on the
mutual information between \(\mathfrak{M}\) and \(\hat{\mathfrak{M}}\),
measured in bits:
\begin{equation}
I_M^{(\mathrm{bits})}(\mathfrak{M}\!:\!\hat{\mathfrak{M}})
\geq
H(\mathfrak{M})
-
h_2(P_{\mathrm{error}})
-
P_{\mathrm{error}}\log_2(|\mathcal{M}|-1),
\label{fano}
\end{equation}
where \(H(\mathfrak{M})\) is the Shannon entropy of the message distribution,
measured in bits, and \(h_2(P_{\mathrm{error}})\) is the binary entropy. This
inequality quantifies how decoding errors reduce the recoverable information.

The mutual information recovered by the decoder cannot exceed the accessible
classical information encoded in the outgoing radiation. Therefore, using the
finite-process area-cost law in Eq.~\eqref{area_budget_bits}, we obtain
\begin{equation}
I_M^{(\mathrm{bits})}(\mathfrak{M}\!:\!\hat{\mathfrak{M}})
\leq
I_{\mathrm{class}}
\lesssim
\frac{1}{4\ln(2)\,\ell_P^2}
\,|\delta A_{\rm rad}|.
\label{area_budget_upper}
\end{equation}

Combining Eqs.~\eqref{fano} and \eqref{area_budget_upper} leads to the
reliability-aware area constraint
\begin{equation}
|\delta A_{\rm rad}|
\gtrsim
4\ln(2)\,\ell_P^2
\max\!\left\{
0,\,
H(\mathfrak{M})
-
h_2(P_{\mathrm{error}})
-
P_{\mathrm{error}}\log_2(|\mathcal{M}|-1)
\right\}.
\label{fano_area}
\end{equation}
The positive part is necessary because the Fano lower bound can become
negative for sufficiently large decoding error probability, whereas the
required area budget is nonnegative.

In the limit of perfect reliability, \(P_{\mathrm{error}}\to0\), and for a
message entropy \(H(\mathfrak{M})=B\) bits, Eq.~\eqref{fano_area} reduces to
\begin{equation}
|\delta A_{\rm rad}|
\gtrsim
4\ln(2)\,\ell_P^2\,B,
\label{per_bit_area}
\end{equation}
showing that, in this regime, each reliably transmitted bit requires a
horizon-area reduction of at least \(4\ln(2)\ell_P^2\).

\section{Fisher-area quantum speed limits and information geometry for the horizon}
\label{sec:fisher_area}

The preceding sections show that, in the near-steady saturated regime, the
entropy and information carried by the radiation field are constrained by the
finite horizon-area budget associated with the emitted scalar quanta. We now
formulate a complementary dynamical constraint in terms of information
geometry. Quantum speed limits provide lower bounds on the time required to
perform a prescribed quantum process and have been formulated for both closed
and open quantum dynamics, including settings in which quantum correlations
modify the accessible dynamical speed
\cite{taddei2013quantum,maleki2020speed,maleki2024universal}.

The construction developed here is based on the temporal Fisher information
of the radiation-field occupation-number distribution. Because the reduced
radiation-field density matrix remains diagonal in a fixed occupation-number
basis, its evolution defines a commuting trajectory of density matrices. Along
this trajectory, the temporal Fisher information of the occupation
probabilities coincides with the corresponding temporal quantum Fisher
information. The duration bound derived below may therefore be interpreted as an
information-geometric quantum speed limit for the occupation-diagonal reduced
dynamics. 

In the diagonal occupation-number description established above, let
\(x\equiv\{n\}\) label a radiation-field configuration, with instantaneous
probability \(p_x(t)\equiv\rho_{\mathrm{diag}}(\{n\},t)\). The Fisher-geometric
expressions below are understood on the fixed support of the probability
distribution over the interval considered.
For the emission process, the initial time
\(t_0\) may be chosen after the onset of the radiative evolution, while the
initial radiation entropy and radiation--environment correlations remain
negligible at the order retained.

The surprisal associated with configuration \(x\) is the stochastic
variable
\begin{equation}
    \mathcal I_x(t)
    :=
    -\ln p_x(t).
    \label{surprisal_def}
\end{equation}
Its expectation value gives the dimensionless entropy of the diagonal
radiation-field density matrix,
\(\langle\mathcal I\rangle=S^{(\rm rad)}\), and therefore
\(S_{\rm phys}^{(\rm rad)}=k_B\langle\mathcal I\rangle\).

The instantaneous standard deviations of the surprisal,
\(\Delta\mathcal I(t)\), and of its time derivative,
\(\Delta\dot{\mathcal I}(t)\), characterize the stochastic variability of
information in the radiative sector. Since probability normalization implies
\(\langle\dot{\mathcal I}\rangle=0\), the Fisher information of the
time-parametrized family \(p_x(t)\) can be written as
\begin{equation}
    I_F(t)
    =
    \sum_x p_x(t)
    \left[\partial_t\ln p_x(t)\right]^2
    =
    \sum_x\frac{\dot p_x(t)^2}{p_x(t)}
    =
    \big[\Delta\dot{\mathcal I}(t)\big]^2.
    \label{time_fisher_def}
\end{equation}
This dynamical, or temporal, Fisher information measures the instantaneous
speed of the occupation-number distribution along its trajectory in
statistical space. Formally, it is the Fisher information of the
time-parametrized family \(p_x(t)\), with the physical evolution time \(t\)
serving as the parameter~\cite{ito2020stochastic}. Unlike the usual
estimation-theoretic setting, however, \(t\) is not an externally encoded
control parameter whose value is to be inferred from measurement outcomes; it
labels the actual dynamical evolution of the reduced radiation-field density
matrix
\cite{Fisher1925,Rao1945,Cramer1946,maleki2021quantum2}.
The square root of the Fisher information of the time corresponds to quantum speed limit of the density matrix
\cite{taddei2013quantum,maleki2024universal}.


The thermodynamic time-information uncertainty relation for the entropy rate
gives~\cite{nicholson2020time}
\begin{equation}
    \frac{|\dot S_{\rm phys}^{(\rm rad)}(t)|}{k_B}
    =
    |\dot S^{(\rm rad)}(t)|
    \leq
    \Delta\dot{\mathcal I}(t)\,
    \Delta\mathcal I(t)
    =
    \sqrt{I_F(t)}\,
    \Delta\mathcal I(t).
    \label{sphysdelti1}
\end{equation}

Using the near-steady dimensionless entropy-area relation in
Eq.~\eqref{Srate_landauer}, Eq.~\eqref{sphysdelti1} gives, for nonvanishing
surprisal fluctuations,
\begin{equation}
    I_F(t)
    \gtrsim
    \left(
        \frac{|\dot A_{\rm rad}(t)|}
        {4\ell_P^2\Delta\mathcal I(t)}
    \right)^{\!2}.
    \label{Fisher_area_rate}
\end{equation}
Thus, sustaining a given radiative horizon-area rate requires a minimum
Fisher information of the occupation-number distribution and, equivalently, a
minimum speed of the occupation-diagonal reduced density matrix.

The same constraint admits an equivalent energetic representation. Using the
near-steady relation
\(|\dot S^{(\rm rad)}(t)|
\simeq|\dot{\tilde E}_{\rm rad}(t)|/(k_BT_H)\) from
Eq.~\eqref{Srate_landauer}, we obtain
\begin{equation}
    I_F(t)
    \gtrsim
    \left(
        \frac{|\dot{\tilde E}_{\rm rad}(t)|}
        {k_BT_H\Delta\mathcal I(t)}
    \right)^{\!2}.
    \label{Fisher_energy_bound}
\end{equation}
Equation~\eqref{Fisher_energy_bound} is the energetic representation of the
same information-geometric constraint, rather than an independent bound,
because the corotating energy flux and the radiative horizon-area rate are
related through Eq.~\eqref{Srate_landauer}.

If the protocol constrains the surprisal fluctuations according to
\(\Delta\mathcal I(t)\leq\Delta\mathcal I_{\max}\), then
Eq.~\eqref{Fisher_area_rate} implies
\begin{equation}
    I_F(t)
    \gtrsim
    \left(
        \frac{1}
        {4\ell_P^2\Delta\mathcal I_{\max}}
    \right)^{\!2}
    |\dot A_{\rm rad}(t)|^2.
    \label{Fisher_area_rate_max}
\end{equation}
Therefore, rapid changes in the radiative horizon-area budget require a
correspondingly large temporal Fisher information. Equivalently, a radiation
process associated with a large horizon-area or corotating-energy flux must
move rapidly through the information geometry of the reduced density-matrix
trajectory.

It is also useful to express this statement in integrated form. The Fisher
length accumulated over a time interval \([t_0,t_1]\) is defined as
\begin{equation}
    \ell_F
    :=
    \int_{t_0}^{t_1}\sqrt{I_F(t)}\,dt
    =
    \int_{t_0}^{t_1}\Delta\dot{\mathcal I}(t)\,dt.
    \label{Fisher_length_def}
\end{equation}
This quantity measures the Fisher--Rao length traversed by the evolving
occupation-number distribution~\cite{Amari2016}. Since
\(I_F(t)=\mathcal F_Q(t)\) for the commuting density-matrix family considered
here, \(\ell_F\) is exactly twice the corresponding Bures path length,
\(\ell_{\rm Bures}=\ell_F/2\). It is therefore a natural integrated measure of
the quantum-information-geometric speed entering the correlation-generation
speed limit derived below.

Using Eq.~\eqref{Fisher_area_rate} together with
\(\Delta\mathcal I(t)\leq\Delta\mathcal I_{\max}\), and noting that
\(\int_{t_0}^{t_1}|\dot A_{\rm rad}(t)|\,dt
\geq|\delta A_{\rm rad}|\), we obtain
\begin{equation}
    \ell_F
    \gtrsim
    \frac{1}{4\ell_P^2\Delta\mathcal I_{\max}}
    \int_{t_0}^{t_1}
    |\dot A_{\rm rad}(t)|\,dt
    \geq
    \frac{|\delta A_{\rm rad}|}
    {4\ell_P^2\Delta\mathcal I_{\max}}.
    \label{Fisher_area_budget}
\end{equation}
Thus, the radiative area change imposes a lower bound on the
information-geometric distance traversed by the reduced radiation-field
density matrix.

For the recoverable classical information carried by the radiation channel,
we use the finite-process entropy-to-information conversion introduced in
Sec.~\ref{sec:bits_area}. Assuming that the outgoing radiation has negligible
initial entropy and carries no initially encoded classical information, the
Holevo bound gives
\begin{equation}
    I_{\mathrm{class}}[t_0\!\to\!t_1]
    \lesssim
    \frac{\Delta S^{(\rm rad)}}{\ln(2)}.
    \label{Fisher_Holevo_finite}
\end{equation}
The radiative entropy change over the same interval satisfies
\begin{align}
    \Delta S^{(\rm rad)}
    &\leq
    \int_{t_0}^{t_1}
    |\dot S^{(\rm rad)}(t)|\,dt
    \nonumber\\
    &\leq
    \int_{t_0}^{t_1}
    \Delta\mathcal I(t)\sqrt{I_F(t)}\,dt
    \nonumber\\
    &\leq
    \Delta\mathcal I_{\max}\,\ell_F.
    \label{Fisher_entropy_length}
\end{align}
Combining Eqs.~\eqref{Fisher_Holevo_finite} and
\eqref{Fisher_entropy_length} yields
\begin{equation}
    I_{\mathrm{class}}[t_0\!\to\!t_1]
    \lesssim
    \frac{\Delta\mathcal I_{\max}}{\ln(2)}\,\ell_F.
    \label{Fisher_information_tradeoff}
\end{equation}
Thus, a finite Fisher length limits the amount of recoverable classical
information that can be transmitted through the radiation channel.

The same reasoning also constrains correlation generation. Using the
Araki--Lieb bound
\(I_M(\mathcal R\!:\!\mathcal E)\leq2S(\mathcal R)\), and assuming that both
the initial radiation--environment mutual information and the initial
outgoing-radiation entropy are negligible, we obtain
\begin{align}
    \Delta I_M(\mathcal R\!:\!\mathcal E)
    &\simeq
    I_M(\mathcal R\!:\!\mathcal E;t_1)
    \nonumber\\
    &\leq
    2S(\mathcal R;t_1)
    \simeq
    2\Delta S(\mathcal R)
    \nonumber\\
    &\lesssim
    2\Delta\mathcal I_{\max}\,\ell_F.
    \label{Fisher_MI_length}
\end{align}
Expressed in bits, this becomes
\begin{equation}
    \Delta I_M^{(\mathrm{bits})}(\mathcal R\!:\!\mathcal E)
    \lesssim
    \frac{2\Delta\mathcal I_{\max}}{\ln(2)}\,\ell_F.
    \label{Fisher_MI_bits}
\end{equation}
Therefore, generating \(\mathcal B\) bits of radiation--environment mutual
information requires
\begin{equation}
    \ell_F
    \gtrsim
    \frac{\mathcal B\ln(2)}
    {2\Delta\mathcal I_{\max}}.
    \label{Fisher_length_bits}
\end{equation}

The area- and energy-resolved Fisher bounds in
Eqs.~\eqref{Fisher_area_rate}--\eqref{Fisher_area_budget} inherit the
near-steady, thermally saturated HBAR relation. By contrast, the
correlation-generation constraint in
Eqs.~\eqref{Fisher_MI_length}--\eqref{Fisher_length_bits} follows directly
from the time-information uncertainty relation and the Araki--Lieb bound. It
therefore does not require an additional use of the entropy-area
correspondence, although it retains the fixed-support,
negligible-initial-correlation, and bounded-surprisal-fluctuation assumptions
stated above.

Let \(\mathcal T=t_1-t_0\) denote the duration of the protocol, and define the
time-averaged temporal Fisher information by
\begin{equation}
    \overline I_F
    :=
    \frac{1}{\mathcal T}
    \int_{t_0}^{t_1}I_F(t)\,dt.
    \label{average_Fisher_def}
\end{equation}
The Cauchy--Schwarz inequality gives
\begin{equation}
    \ell_F
    =
    \int_{t_0}^{t_1}\sqrt{I_F(t)}\,dt
    \leq
    \mathcal T\sqrt{\overline I_F}.
    \label{Fisher_CS}
\end{equation}
Combining Eqs.~\eqref{Fisher_length_bits} and \eqref{Fisher_CS} yields
\begin{equation}
    \mathcal T
    \gtrsim
    \frac{\mathcal B\ln(2)}
    {2\Delta\mathcal I_{\max}\sqrt{\overline I_F}}.
    \label{Tmin_speedlimit}
\end{equation}
Equation~\eqref{Tmin_speedlimit} is the Fisher-information counterpart of the
area-cost laws derived above and constitutes an information-geometric 
speed limit time for correlation generation in the occupation-diagonal reduced
dynamics. 
A finite value of this speed
therefore imposes a lower bound on the duration required to generate a
prescribed amount of radiation--environment mutual information.

Beyond this dynamical constraint, the finite horizon-area budget established
above also constrains how the correlations generated by the HBAR process may
be shared once established. This restriction extends beyond the bipartite
radiation--environment description to multipartite settings through
bipartitions of the degrees of freedom complementary to the outgoing
radiation. A general formulation of this area-bounded sharing of correlations
is presented in Appendix~\ref{sec:area_monogamy}.
\section{Conclusions}

In this work, we have developed a geometric information-theoretic framework
for the Horizon-brightened acceleration radiation (HBAR) channel, in which the
radiative contribution to the horizon-area change provides the geometric
budget underlying information processing. Starting from the quantum-optical
description of atom--field interactions in Kerr geometry, we used the thermal
radiation-field density matrix and its formal correspondence with black-hole
thermodynamics to derive a set of area--information trade-offs.

The thermodynamic foundation of these results was formulated through Spohn's
theorem. In general, the radiation-field evolution is accompanied by a
nonnegative irreversible entropy production. In the near-steady, thermally
saturated regime, this balance reduces to the relation connecting the
radiation entropy, the corotating energy carried by the emitted scalar
quanta, and the associated radiative horizon-area change. The area-cost laws
derived in this work apply within this regime.

For accessible classical information, the Holevo bound yields a
leading-order lower bound of \(4\ln(2)\,\ell_P^2\) on the radiative
horizon-area cost per bit. Fano's inequality incorporates a prescribed
decoding error probability into this geometric cost. From the perspective of
correlations, the mutual information between the radiation field and its
environment is bounded by twice the radiation entropy, yielding a
corresponding lower bound of \(2\ln(2)\,\ell_P^2\) on the area cost per bit of
total mutual information. The difference between these coefficients reflects the
distinction between classically accessible information and the total
correlations supported by a bipartite quantum system. The appendix further
shows that the same entropy budget constrains the sharing of correlations
between the radiation and complementary partitions of its environment.

We also derived dynamical constraints from the Fisher information of the
radiation-field occupation-number distribution. The resulting bounds relate
the statistical speed of the radiative evolution to the horizon-area and
corotating-energy fluxes. Their integrated form constrains the Fisher length
required by a finite area budget and yields a corresponding lower bound on
the duration required to generate a prescribed amount of correlation. These relations provide an
information-geometric characterization of the dynamical cost associated with
the HBAR process.

Together, the entropy, area, and information bounds establish a coherent
bits-per-area principle for the HBAR channel. Their origin lies in the
near-horizon detailed-balance structure that yields the thermal
radiation-field density matrix and, in the saturated regime, connects its
entropy and corotating energy to the radiative horizon-area change. The
resulting framework therefore brings black-hole thermodynamics, quantum
information, and information geometry into a common geometric description,
without relying on a particular encoding of the classical information.

Although the HBAR setup remains theoretical, its underlying near-horizon
conformal quantum-mechanical structure suggests that the geometric
information-theoretic framework developed here may extend to other horizon
geometries and related acceleration-radiation settings. Analogue systems that
simulate relative acceleration
\cite{Fulling-Wilson_2018_EP,Ben-Benjamin-etal_2019_Unruh-rev}, including
dynamical-Casimir configurations in microwave cavities
\cite{Svidzinsky_2018_mirrors-virtual}, may provide useful settings in which
aspects of this framework could eventually be explored. Establishing such a
connection, however, requires a dedicated operational model relating the
radiative area budget to experimentally accessible observables.

\begin{acknowledgments}
This work was partially supported by the U.S. Department of Energy
(DE-SC-0023103, DE-SC0024882); Department of Energy
Contract (DE-AC36-08GO28308, SUB-2023-10388); and the Welch
Foundation (A-1261). It was also partially supported by the Air Force Office
of Scientific Research under Grant No. FA9550-21-1-0017 (C.R.O. and G.V.-M.).
C.R.O. and G.V.-M. were partially supported by the Army Research Office (ARO)
under Grant No. W911NF-23-1-0202. G.V.-M. gratefully acknowledges the Center
for Mexican American and Latino/a Studies at the University of Houston for
its generous support through a Lydia Mendoza Fellowship. H.E.C. acknowledges
support from the University of San Francisco Faculty Development Fund.
\end{acknowledgments}

\appendix
\section{Area-bounded sharing of correlations}
\label{sec:area_monogamy}

Let \(\mathcal R\) denote the outgoing acceleration radiation, and let the
complementary sector \(\mathcal E\) introduced in
Sec.~\ref{sec:mi_area} be partitioned into two disjoint registers,
\(\mathcal E_1\) and \(\mathcal E_2\). For instance,
\(\mathcal E_1\) may contain the black-hole and atomic degrees of freedom,
while \(\mathcal E_2\) may represent an external memory or degrees of freedom
accessible to a distant observer. We define the von~Neumann mutual
informations
\begin{equation}
\begin{aligned}
    I_M(\mathcal R\!:\!\mathcal E_1)
    &=
    S(\mathcal R)
    +
    S(\mathcal E_1)
    -
    S(\mathcal R\mathcal E_1),
    \\
    I_M(\mathcal R\!:\!\mathcal E_2)
    &=
    S(\mathcal R)
    +
    S(\mathcal E_2)
    -
    S(\mathcal R\mathcal E_2),
\end{aligned}
\label{MI_defs}
\end{equation}
where \(S(\cdot)\) denotes the dimensionless von~Neumann entropy in nats.

For a tripartite density matrix
\(\hat\rho_{\mathcal R\mathcal E_1\mathcal E_2}\), the following general,
model-independent inequality holds:
\begin{equation}
    I_M(\mathcal R\!:\!\mathcal E_1)
    +
    I_M(\mathcal R\!:\!\mathcal E_2)
    \leq
    2S(\mathcal R).
    \label{MI_mono_core}
\end{equation}

\paragraph{Proof from weak monotonicity.}

The left-hand side of Eq.~\eqref{MI_mono_core} can be written as
\begin{align}
    I_M(\mathcal R\!:\!\mathcal E_1)
    +
    I_M(\mathcal R\!:\!\mathcal E_2)
    &=
    2S(\mathcal R)
    +
    S(\mathcal E_1)
    +
    S(\mathcal E_2)
    \nonumber\\
    &\quad
    -
    S(\mathcal R\mathcal E_1)
    -
    S(\mathcal R\mathcal E_2).
    \label{MI_sum_D}
\end{align}
Weak monotonicity, equivalently strong subadditivity of the von~Neumann
entropy~\cite{lieb1973proof}, states that
\begin{equation}
    S(\mathcal R\mathcal E_1)
    +
    S(\mathcal R\mathcal E_2)
    \geq
    S(\mathcal E_1)
    +
    S(\mathcal E_2).
    \label{weak_monotonicity}
\end{equation}
Substitution of Eq.~\eqref{weak_monotonicity} into
Eq.~\eqref{MI_sum_D} proves Eq.~\eqref{MI_mono_core}.

\paragraph{Area form.}

In the near-steady saturated regime, the entropy-area relation gives
\begin{equation}
    S(\mathcal R;t_1)-S(\mathcal R;t_0)
    \simeq
    \frac{1}{4\ell_P^2}
    \int_{t_0}^{t_1}
    |\dot A_{\rm rad}(t)|\,dt.
    \label{sharing_entropy_area}
\end{equation}
If the outgoing radiation is initially prepared with negligible entropy and
negligible correlations, and if the radiative area change has a definite sign
over the interval, then
\(S(\mathcal R;t_1)\simeq|\delta A_{\rm rad}|/(4\ell_P^2)\). Combining this
relation with Eq.~\eqref{MI_mono_core} gives
\begin{equation}
    \Delta I_M(\mathcal R\!:\!\mathcal E_1)
    +
    \Delta I_M(\mathcal R\!:\!\mathcal E_2)
    \lesssim
    \frac{|\delta A_{\rm rad}|}{2\ell_P^2}
    =
    \frac{c^3}{2\hbar G}
    |\delta A_{\rm rad}|.
    \label{MI_mono_area}
\end{equation}
The mutual information in Eq.~\eqref{MI_mono_area} is measured in nats.
Expressed in bits, the right-hand side is divided by \(\ln(2)\).

\paragraph{Beyond two parties: bipartitioned bound.}

Suppose that the complementary sector is further partitioned as
\(\mathcal E=\mathcal E_1\mathcal E_2\cdots\mathcal E_m\). For any
bipartition of these registers into two disjoint groups
\(\mathcal G\) and \(\overline{\mathcal G}\), weak monotonicity gives
\begin{equation}
    I_M(\mathcal R\!:\!\mathcal G)
    +
    I_M(\mathcal R\!:\!\overline{\mathcal G})
    \leq
    2S(\mathcal R).
    \label{bipartition_mono}
\end{equation}
Thus, each bipartition defines a trade-off between the correlations that the
two complementary groups may share with the outgoing radiation, with the total
budget bounded by \(2S(\mathcal R)\) and, consequently, by the available
radiative area budget.

Equation~\eqref{bipartition_mono} is a bipartitioned sharing constraint; it
does not imply the fully additive inequality
\(\sum_i I_M(\mathcal R\!:\!\mathcal E_i)\leq2S(\mathcal R)\).
Rather, it provides a family of area-bounded trade-offs, one for each chosen
division of the complementary degrees of freedom. In a communication
architecture, the relevant bipartition may separate a preferred receiver from
all remaining registers.

\bibliography{apssamp}

\end{document}